\documentclass[english]{epl}
\usepackage{babel}
\usepackage[T1]{fontenc}
\usepackage[latin1]{inputenc}
\usepackage{graphicx}
\usepackage{latexsym}

\title{The  hyperon neutron star mean field model}
\author{Ilona Bednarek \inst{1}, Ryszard Ma\'{n}ka \inst{1}}

\institute{
\inst{1} Department of Astrophysics and Cosmology, Institute of Physics, \\
  University of Silesia, Uniwersytecka 4, PL-40-007 Katowice, Poland.
  }

\begin{document}

\maketitle

\begin{abstract}
In this paper the impact of the strength of hyperon-nucleon and
hyperon-hyperon on the maximum neutron star mass is shown.
The strong hyperon-hyperon
coupling constant together with the additional nonlinear meson
interaction terms lead to the essential softening of the equation
of state and to the existence of another more compact branch of
stable neutron star configurations. The analysis of the
mass-radius relations leads to the conclusion that the additional
stable configurations of hyperon-rich neutron stars are
characterized by the reduced value of radii.
\end{abstract}

\begin{figure}
\begin{center}
\onefigure{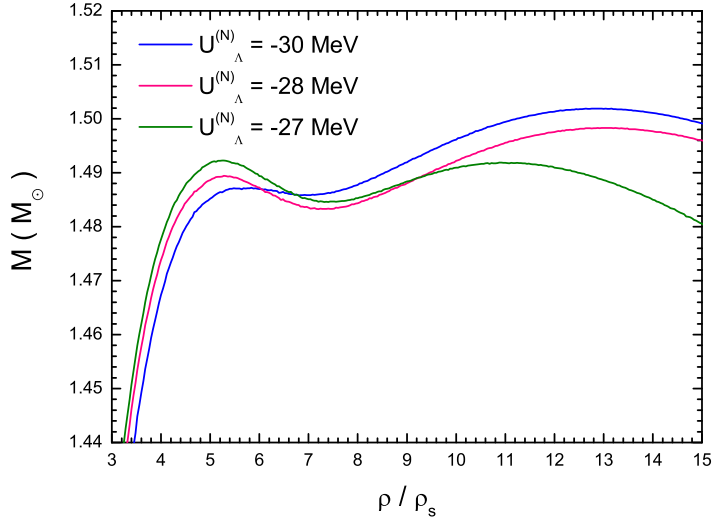}
\caption{(color online) The mass-central density relation. The left panel shows
results for the ordinary TM1 parameter set whereas the right panel
presents the mass-radius relation for the TM$1^{*}$ parameter set.
} \label{fig:mrhotm1s}
\end{center}
\end{figure}

\begin{figure}
\begin{center}
\onefigure{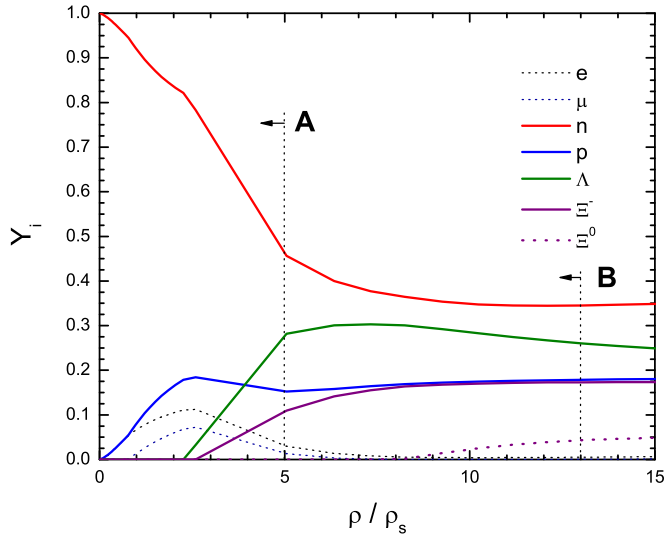}
\caption{(color online) Baryon and lepton concentrations in neutron star matter
as a function of density $\rho$ ($\varepsilon =\rho c^2$ ) where
$\varepsilon$ denotes the energy density.} \label{fig:Yrhotm1s}
\end{center}
\end{figure}

\begin{figure}
\begin{center}
\onefigure{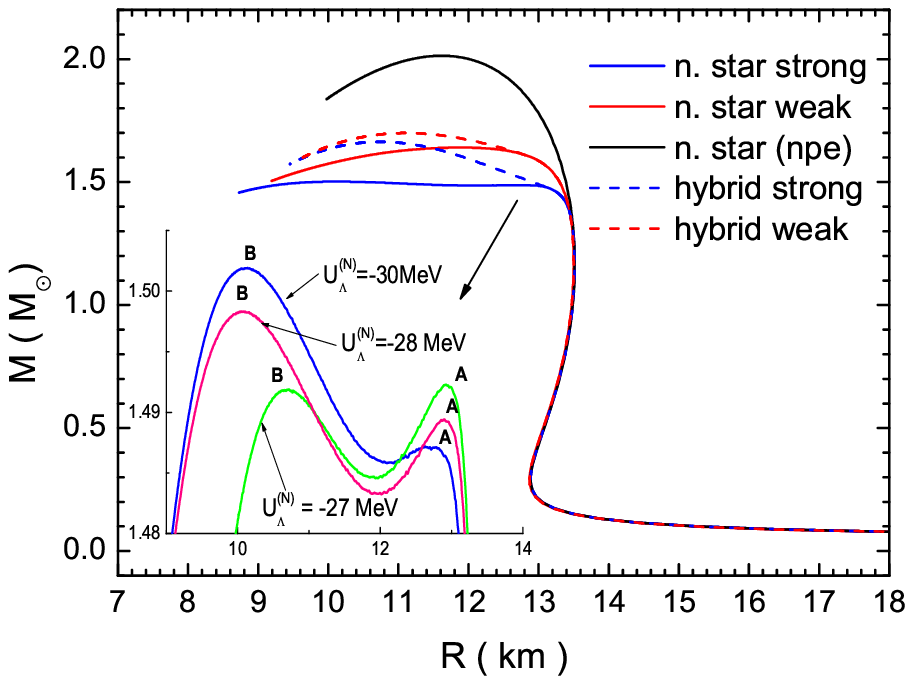}
\caption{(color online) The mass-radius relation for the TM1*  parameter sets.} \label{fig:rmtm1s}
\end{center}
\end{figure}

The theoretical framework  employed in this
paper is the relativistic nuclear mean field theory extended by
allowing for additional non-linear meson couplings.   The chiral
effective Lagrangian proposed by Furnstahl, Serot and Tang (FST)
\cite{FST1}, \cite{FST2} constructed on the basis of the effective
field theory and density functional theory for hadrons gave in the
result the extension of the standard relativistic mean field
theory and introduced additional non-linear scalar-vector and
vector-vector self-interactions. This Lagrangian in general
includes all non-renormalizable couplings consistent with the
underlying symmetries of QCD. Applying the  dimensional analysis
of Georgi and Manohar \cite{georgi1}, \cite{georgi2} and the
concept of naturalness one can expand the nonlinear Lagrangian and
organize it in increasing powers of the fields and their
derivatives and truncated at given level of accuracy \cite{serot}.
If the truncated Lagrangian includes terms up to the forth order
it can be written in the following form
\begin{eqnarray}
\mathcal{L} & = & \sum_{B}
\bar{\psi}_B(i\gamma^{\mu}D_{\mu}-m_B+g_{\sigma
B}\sigma+g_{\sigma^{\ast}B}\sigma^{\ast})\psi_B \\ \nonumber &+&
\frac{1}{2}\partial _{\mu }\sigma \partial ^{\mu }\sigma -m_{\sigma}^2\sigma^2(\frac{1}{2}+\frac{\kappa_3}{3!}\frac{g_{\sigma B}\sigma}{M}+\frac{\kappa_4}{4!}\frac{g_{\sigma B}^2\sigma^2}{M^2})+ \frac{1}{2}\partial _{\mu }\sigma^* \partial ^{\mu }\sigma^*- \frac{1}{2}m^{2}_{\sigma^*}\sigma^{*2} \nonumber \\
 & + &\frac{1}{2}m^{2}_{\phi}\phi_{\mu}\phi ^{\mu} -\frac{1}{4}\phi_{\mu \nu}\phi^{\mu \nu}
 -\frac{1}{4}\Omega _{\mu \nu }\Omega ^{\mu \nu }+\frac{1}{2}(1+\eta_{1}\frac{g_{\sigma B}\sigma}{M}+\eta_2\frac{g_{\sigma B}^2\sigma^2}{2 M^2})m_{\omega }^2\omega _{\mu }\omega ^{\mu } \nonumber \\
 & - & \frac{1}{4}R_{\mu \nu }^{a}R^{a\mu \nu }+(1+\eta_{\rho}\frac{g_{\sigma B}\sigma}{M})\frac{1}{2}m_{\rho} ^2\rho^{a}_{\mu }\rho^{a\mu }+\frac{1}{24}\zeta_0 g_{\omega B}^2(\omega _{\mu }\omega ^{\mu })^{2}. \nonumber
 \label{lag1}
\end{eqnarray}
$\Psi_B^T
=(\psi_N,\psi_{\Lambda},\psi_{\Sigma},\psi_{\Xi})$.  The covariant
derivative $D_{\mu}$ is defined as
\begin{equation}
D_{\mu}=\partial_{\mu}+ig_{\omega B}\omega_{\mu}+ig_{\phi
B}\phi_{\mu}+ig_{\rho B}I_{3B}\tau^a\rho^a_{\mu}
\end{equation}
whereas  \( R_{\mu \nu }^{a} \), \( \Omega _{\mu \nu } \) and \(
\phi_{\mu \nu } \) are the field tensors. $m_B$ denotes baryon
mass whereas \( m_{i} \) $(i= \sigma ,\omega ,\rho ,\sigma^*,\phi
)$ are masses assigned to the meson fields, $M$ is the nucleon
mass.
 In the high density regime in neutron star interiors when the Fermi
energy of nucleons exceeds the hyperon masses hyperons are
expected to emerged due to the strangeness changing interactions
\cite{weber}, \cite{glen}, \cite{glen1}, \cite{bema}, \cite{gal}.
 The appearance of these additional degrees of freedom and their impact
on a neutron star structure have been the subject of extensive
studies. In order to reproduce attractive hyperon-hyperon
interaction two additional hidden-strangeness mesons, which do not
couple  to nucleons, have been introduced, namely the scalar meson
$f_0(975)$
(denoted as $\sigma^{\ast}$) and the vector meson $\phi(1020)$ \cite{Schaffner}.\\
Due to the fact that the expectation value of the $\rho$ meson
field is an order of magnitude smaller than that of $\omega$ meson
field, the Lagrangian function (\ref{lag1}) does not include the
quartic $\rho$ meson term. In addition, as this paper deals with
the problem of infinite nuclear matter the terms in the original
Lagrangian function (see \cite{FST1}, \cite{FST2}) involving
tensor couplings and meson field gradients have been excluded.
 There are two parameter sets presented in the original paper by Furnstahl et al
\cite{FST1}, \cite{FST2} G1 and G2 which have been determined by
calculating nuclear properties such as binding energies, charge
distribution and spin-orbit splitting for a selected set of nuclei
\cite{sil}.
For the purposes of this paper two parameter sets have been chosen: the G2 parameter
set and the extended TM1 parameter set denoted as TM$1^{*}$. The
latter one constructed by Del Estal et al \cite{estal1}
represents the  standard TM1 parameter set supplemented with
additional nonlinear couplings stemming from the effective field
theory. Calculations performed with the TM$1^{*}$ parameter set
properly reproduce properties of finite nuclei. Both parameter
sets G2 and TM$1^{*}$ make it possible to compare the obtained
results with the Dirac-Brueckner-Hartree-Fock (DBHF) calculations
for nuclear and neutron matter above the saturation density. The
DBHF method results in a rather soft equation of state in the
vicinity of the saturation point and  for higher densities.
Calculations performed on the basis of the effective FST
Lagrangian with the use of the G2 and TM$1^{*}$ parameter sets
predict similar, soft equation of state.\\
The theoretical description  of strange hadronic matter properties
is given  within the relativistic mean field approximation. In
this approximation meson fields are separated into classical mean
field values and quantum fluctuations which are not included in
the ground state. In the case of a hyperon-rich  matter  the
considered parameterization has to be supplemented by the
parameter set related to the strength of the hyperon-nucleon and
hyperon-hyperon interactions. The scalar meson coupling to
hyperons can be calculated from the potential depth of a hyperon
in the saturated nuclear matter. 
Assuming the SU(6) symmetry for the vector coupling  and
determining the scalar coupling constants from the
potential depths, the hyperon-meson couplings can be fixed.\\
 The strength of hyperon coupling
to strange meson $\sigma^{\ast}$ is restricted through the
following relation \cite{gal}
$
U^{(\Xi)}_{\Xi}\approx U^{(\Xi)}_{\Lambda}\approx
2U^{(\Lambda)}_{\Xi}\approx 2U^{(\Lambda)}_{\Lambda}.
$
which together with the estimated value of hyperon potential
depths in hyperon matter provides effective constraints  on scalar
coupling constants to the $\sigma^{\ast}$ meson. The currently
obtained value of the $U^{(\Lambda)}_{\Lambda}$ potential at the
level of 5 MeV \cite{Takahashi} permits the existence of
additional parameter set \cite{bema2} which reproduces this weaker
$\Lambda\Lambda$ interaction. Throughout this paper  this
parameter set is referred to as weak, whereas strong denotes the
stronger $\Lambda\Lambda$ interaction for
$U^{(\Lambda)}_{\Lambda}\simeq 20$ MeV \cite{Schaffner}.
\begin{table}
\caption{Chosen parameter sets.}\label{tab:TM1a}
\begin{center}
\begin{tabular}{lllllllllll}
\hline
&$m_{\sigma}$&$g_{\sigma
N}$&$g_{\omega N}$&$g_{\rho N}$&$\eta_1$&$\eta_2$&$\eta_{\rho}$&$\zeta_0$&$\kappa_3$&$\kappa_4$\\
\hline
G2&520.25&10.496&12.76&9.48&0.65&0.11&0.39&2.64&3.247&0.63\\\hline
TM$1^{*}$&511.20&11.22&14.98&10.00&1.1&0.1&0.45&3.6&2.513&8.97\\\hline
\end{tabular}
\end{center}
\end{table}
\begin{table}
\caption{Strange scalar sector parameters.}
\label{tab:sscalar}
\begin{center}
\begin{tabular}{ll|l|l|l|l}
\hline
&&$g_{\sigma\Lambda}$&$g_{\sigma\Xi}$&$g_{\sigma^{\ast}\Lambda}$&$g_{\sigma^{\ast}\Xi}$\\
\hline G2&weak&6.410&3.337&3.890&11.643\\\cline{2-6} &strong&6.410&3.337&7.931&12.458 \\
\hline
&&$g_{\sigma\Lambda}$&$g_{\sigma\Xi}$&$g_{\sigma^{\ast}\Lambda}$&$g_{\sigma^{\ast}\Xi}$\\
\hline TM$^{*}$&weak&6.971&3.583&5.526&13.450\\\cline{2-6} strong&6.971&3.583&9.069&14.394\\
\hline
\end{tabular}
\end{center}
\end{table}
In the nucleon sector the saturation properties of nuclear matter
namely the saturation density $\rho_s$, binding energy $E_b$,
compressibility $K$ and  Dirac effective nucleon mass $m_{eff
N}=M-g_{\sigma N}\sigma$ determined for the given parameter sets.
Neutron star matter is considered as a system with conserved
baryon number $n_b=\sum_B n_B$ ($n_B=k_B^3/3\pi^2$ denotes the
number density of species $B$, $B=n,p,\Lambda,\Xi^-, \Xi^0$) and
electric charge being in chemical equilibrium with respect to weak
decays. In general, weak processes for baryons can be written in
the following form $B_1+l\leftrightarrow B_2$ where $B_1$ and
$B_2$ are baryons, $l$   denotes lepton. Provided that the  weak
processes stated above  take place in thermodynamic equilibrium
the following relation between chemical potentials can be
established $\mu_B=q_B\mu_n-q_{e_{B}}\mu_e$. This relation
involves two independent chemical potentials $\mu_n$ and $\mu_e$
corresponding to baryon number and electric charge conservation,
$\mu_B$ denotes chemical potential of baryon $B$ with the baryon
number $q_B$ and the electric charge $q_{e_{B}}$.
The baryon effective mass is defined as $m_{eff B}=m_B-g_{\sigma
B}\sigma_0-g_{\sigma^{*}B}\sigma^{*}_0$. Thus the $\beta$
equilibrium conditions for $\Lambda$, $\Sigma$ and $\Xi$ hyperons
lead to the following relations:
\begin{equation}
\mu_{\Lambda}=\mu_{\Sigma^{0}}=\mu_{\Xi^{0}}=\mu_n,
\hspace{5mm}\mu_{\Sigma^{-}}=\mu_{\Xi_{-}}=\mu_n+\mu_e,
\hspace{5mm}\mu_p=\mu_{\Sigma^{+}}=\mu_{n}-\mu_{e}.
\end{equation}
The conditions of charge neutrality and
 $\beta$-equilibrium
imply the presence of leptons which are introduced as free
particles. 
Muons start to appear in neutron star matter in the process
$e^-\leftrightarrow \mu^-$ after $\mu_{\mu}$ has reached the value
equal to the muon mass. The appearance of muons not only reduces
the number of  electrons but also affects the proton fraction.
\newline
Solving the hydrostatic equilibrium equation of Tolman,
Oppenheimer and Volkov, which allows one to construct  theoretical
model of a neutron star and to specify and analyze its properties,
demands the specification of the equation of state.
\begin{figure}
\begin{center}
\onefigure{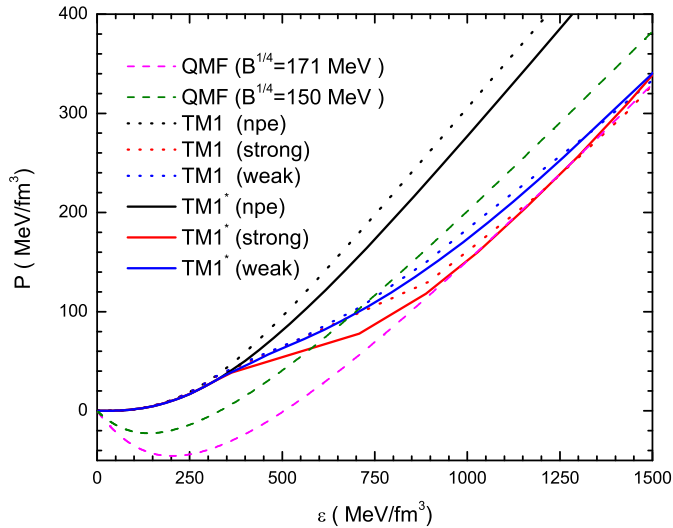}
\caption{(color online) The equation of state calculated for TM1 and TM$1^{*}$
parameter sets. Dotted curves represent the results obtained for
TM1 parameter set. The influence of hyperons is considered for the
weak and strong $Y-Y$ interactions. The case of non-strange
baryonic  matter is marked by npe. Dashed curves show the quark
matter equations of state obtained for two fixed values of the bag
parameter: $B^{1/4}= 150$ MeV and $B^{1/4}= 171$ MeV ($B_{crit}$)}
\label{fig:eostm1s}
\end{center}
\end{figure}
In Fig.\ref{fig:eostm1s} the pressure of a neutron star matter as
a function of the energy density for the TM1 and TM$1^*$ parameter
sets has been presented. This figure includes results for the
strong and weak $Y-Y$ interactions. For comparison the equations
of state for the matter with zero strangeness has been included.
Dashed curves represent the equations of state obtained  for
neutron star matter without hyperons. The parameter set TM$1^{*}$
leads to more soft equation of state than the TM1 one. This
softening is caused by the presence of  additional nonlinear
couplings.
 In general the inclusion of hyperons softens considerably the
equation of state at high densities. This effect is maximized in
the case of strong $Y-Y$ interaction and by adding nonlinear
terms. The form of the equation of state has profound consequences
for the maximum mass of a neutron star.
 Solutions of the Oppenheimer-Tolman-Volkov equation for the
considered equations of state are presented in Fig.
\ref{fig:rmtm1s}. The obtained mass-radius relations are
constructed for neutron star matter with hyperons and compared
with the mass-radius relation for non-strange matter. The analysis
has been done for the ordinary TM1 parameter set  \cite{Sugahara}
and for TM$1^{*}$. The TM1
parameter set gives larger masses than the TM$1^{*}$ one. However,
the key difference between the TM1 and TM$1{*}$ mass-radius
diagrams lies in the results obtained for the strong
hyperon-hyperon interaction. In this case for TM$1^{*}$ parameter
set besides the ordinary stable neutron star branch there exists
the additional stable branch of solutions which are characterized
by a similar value of masses but with significantly reduced radii.
For the purpose of this paper A denotes the maximum mass
configuration of the ordinary neutron star branch whereas B the
additional maximum. The comparison of the maximum mass
configurations obtained for the weak and strong $Y-Y$ interactions
makes it possible to estimate the role of the
 hyperon-hyperon couplings strength. The strong model
gives the reduced value of the maximum mass. The reduction is of
the order of 0.1-0.2 $M_{\odot}$. The influence of the strength of
the hyperon-nucleon couplings  has been analyzed for a very
limited range of the $g_{\sigma \Lambda}$ parameter strictly
connected with the value of the potential felt by a single
$\Lambda$ in saturated nuclear matter. The chosen values of the
potential are: -27 MeV, -28 MeV, -30 MeV. This leads to the
following $g_{\sigma \Lambda}$ parameters: 9.203, 9.158, 9.069. As
it was stated, the additional nonlinear meson-meson interaction
terms and the strong hyperon-hyperon interaction create necessary
conditions for the existence of the second branch of stable
neutron star configurations. For the ordinary neutron star
configurations the well known results has been obtained i.e.the
weaker the coupling $g_{\sigma \Lambda}$ the lower is the value of
the maximum mass. However, for the additional branch the analysis
of the mass radius relation Fig.\ref{fig:rmtm1s} gives the
opposite result. Similar conclusions can be drown from
Fig.\ref{fig:mrhotm1s}. In this figure the mass as a function of
the central density is depicted. Particular curves are labelled by
the three chosen values of the $U_{\Lambda}^{(N)}$ potential.
The composition of hyperon star matter as well as the  threshold
density for hyperons  are altered when the hyperon-hyperon
interaction strength is changed. Fig. \ref{fig:Yrhotm1s} presents
fractions of particular baryon species $Y_B$ as a function of the
density $\rho$ for the  strong model of the TM$1^{*}$
parameterization.
At very low density neutrons and protons are the most abundant
baryons. The first strange baryon that emerges  is  $\Lambda$ and
is followed by $\Xi^-$ and $\Xi^0$. The maximum mass configuration
$A$ resembles an ordinary neutron star predominantly composed of
nucleons. Hyperons ($\Lambda$ and $\Xi^0$) appear in the innermost
part of the star. The appearance of the new branch on of stable
solutions on the mass-radius diagram makes it possible to exist
more dense stars with increased concentrations of hyperons. The
appearance of $\Xi^-$ hyperons at higher densities through the
condition of charge neutrality affects the lepton fraction and
causes a drop in their contents. Thus the presence of charged
hyperons permits lower lepton content and in the case of the
additional stable branch objects charge neutrality tends to be
guaranteed without lepton contribution.
The properties of strange neutron stars has been studied with the
use of the improved TM1 parameter set which include additional
nonlinear coupling stemming from the effective field theory.  This
parameter set in the strange sector has been supplemented by
parameters that are related to the strength of the of
hyperon-hyperon and hyperon-nucleon interactions. The impact of
the strength of hyperon interactions on neutron star masses has
been analyzed. This analysis shows that there exists very strong
correlation between the value of the maximum neutron star mass and
the strength of hyperon coupling constants. The inclusion of
additional nonlinear meson interaction terms which modify the high
density behavior of the equation of state together with the strong
hyperon-hyperon interaction lead to the existence of  additional
stable stellar configurations with similar masses and smaller
radii than an ordinary neutron star. The reduction in radius is of
the order of 2.5 km. The internal composition of this additional
neutron star configurations is almost completely free of leptons.
Transmutation similar to the phase transition from the ordinary neutron star
to the more compact hyperstar may be the main origin of the short gamma ray burnst \cite{dar}.\\
In order to complete the analysis of the existence of the very
compact, hyperon rich stars (hyperstar), the conditions under
which   the occurrence of quark matter and the formation of stable
configurations of hybrid stars   have to be
established.
Two phases of matter have been compared: the strange
hadronic matter and quark matter. The phase with the highest
pressure (lowest free energy) is favored. In Fig.\ref{fig:eostm1s}
the equation of state of the quark matter obtained with the use of
the quark mean field model \cite{rm3} with the direct coupling of
the bag parameter to the scalar meson fields $\sigma_0$ and
$\sigma_0^{*}$, is also included. The results strongly depend on
the value of the bag parameter. There exists the limiting value of
$B^{1/4}_{crit}=171$ MeV, for $B > B_{crit}$ there is no quark
phase in the interior of a neutron star. The bag parameters $B <
B_{crit}$ lead to the intersection of the quark matter equation of
state and that of hyperon star matter with the strong $Y-Y$
interactions. Thus a hybrid star with a quark phase inside can be
constructed. The obtained mass-radius relation for hybrid stars
involving quark matter is presented in the right panel of
Fig.\ref{fig:rmtm1s} (for $B^{1/4}=150 MeV < B^{1/4}_{crit}$). The
maximum hybrid star mass is of about $1.52 M_{\odot}$ and is
greater than the maximum mass of the very compact additional
stable branch configurations ($B$). The radius of the hybrid star
maximum mass configuration is also greater than that of the
configuration marked as $B$. 


\end{document}